# Security-Constrained Unit Commitment Considering Locational Frequency Stability in Low-Inertia Power Grids

Mingjian Tuo, *Student Member, IEEE* and Xingpeng Li, *Member, IEEE*

*Abstract*—With increasing installation of wind and solar generation, conventional synchronous generators in power systems are gradually displaced resulting in a significant reduction in system inertia. Maintaining system frequency and rate-of-change-of-frequency (RoCoF) within acceptable ranges becomes more critical for the stability of a power system. This paper first study the impact of inter-area oscillations on the system rate-of-change-of-frequency (RoCoF) security; then, the limitations on locational RoCoF accounting for $G-1$ contingency stability are derived. To capture highest locational RoCoF during the oscillation, multiple measurement windows are introduced. By enforcing these frequency related constraints, a location based RoCoF constrained security constrained unit commitment (LRC-SCUC) model is proposed. Furthermore, an effective piecewise linearization (PWL) technique is employed to formulate a RoCoF linearization problem and linearize the nonlinear function representing the location based RoCoF constraints in SCUC. Case studies are carried on IEEE 24-bus system to demonstrate the effectiveness of proposed LRC-SCUC model. The results also show that deploying virtual inertia techniques not only reduces the total cost, but also improves the system market efficiency.

## NOMENCLATURE

*Sets*

| | |
|---|---|
| $G$ | Set of generators. |
| $G_n$ | Set of generators connected to bus $n$. |
| $K$ | Set of lines. |
| $K^+(n)$ | Set of lines with bus $n$ as receiving bus. |
| $K^-(n)$ | Set of lines with bus $n$ as sending bus. |
| $T$ | Set of time periods. |
| $N$ | Set of buses. |
| $NG$ | Set of generator buses. |
| $N_{loc}$ | Set of local generator buses. |
| $N_{n\text{-}loc}$ | Set of non-local generator buses. |

*Indices*

| | |
|---|---|
| $g$ | Generator $g$. |
| $k$ | Line $k$. |
| $t$ | Time $t$. |
| $n$ | Bus $n$. |

*Parameters*

| | |
|---|---|
| $c_g$ | Linear operation cost for generator $g$. |
| $P_g^{min}$ | Minimum output limit of generator $g$. |
| $P_g^{max}$ | Maximum output limit of generator $g$. |
| $P_k^{max}$ | Long-term thermal line limit for line $k$. |
| $b_k$ | Susceptance of line $k$. |
| $D_{n,t}$ | Forecasted demand at bus $n$ in period $t$. |
| $E_{n,t}$ | Forecasted renewable generation at bus $n$ in period $t$. |
| $R_g^{hr}$ | Ramping limit of generator $g$. |
| $R_g^{re}$ | Reserve ramping limit of generator $g$. |
| $H_g$ | Inertia constant of conventional generator $g$. |
| $c_g^{NL}$ | No load cost for generator $g$. |
| $c_g^{SU}$ | Startup cost of generator $g$. |
| $c_g^{RE}$ | Reserve cost of generator $g$. |
| $c_n$ | Nodal cost for additional inertia at bus $n$. |
| $nT$ | Frequency monitoring window. |
| $\Delta t$ | Number of time periods. |
| $M_t^{Total}$ | Virtual inertia budget. |
| $\Omega$ | A big real number. |
| $T_1$ | First frequency measuring point. |
| $T_2$ | Second frequency measuring point. |

*Variables*

| | |
|---|---|
| $P_{g,t}$ | Output of generator $g$ in period $t$. |
| $m_{g,t}$ | Average nodal inertia in period $t$ after loss of generator $g$. |
| $\Delta m_{g,t}$ | Change in average nodal inertia in period $t$ after loss of generation. |
| $m_t$ | Average nodal inertia in period $t$. |
| $r_{g,t}$ | Reserve from generator $g$ in period $t$. |
| $u_{g,t}$ | Commitment status of generator $g$ in period $t$. |
| $v_{g,t}$ | Start-up variable of generator $g$ in period $t$. |
| $M_t$ | System aggregated inertia in period $t$. |
| $M_t^{VI}$ | Virtual inertia in period $t$. |
| $P_{k,t}$ | Flow online $k$ in period $t$. |
| $\theta_{n,t}$ | Phase angle of bus $n$ in period $t$. |
| $\theta_{m,t}$ | Phase angle of bus $m$ in period $t$. |

## I. INTRODUCTION

Renewable energy sources (RES) have become the promising technology to meet the demand for the reduction of the $CO_2$ emissions. Within the next few decades, generation with synchronous generators will gradually be replaced by converter-connected renewable energy sources such as wind and solar generations. As a result of this transition, current power system may ultimately shift towards a 100% converter-based power system [1]. Traditionally, synchronous generators provide inertia through stored kinetic energy in rotating mass which can counteract frequency excursion during disturbances and thus enhance the frequency stability. While RESs are interfaced to the grid through converters which electrically decouples the rotor's inertia from the whole system [2], contributing little synchronous inertia to the whole system, which is

Mingjian Tuo and Xingpeng Li are with the Department of Electrical and Computer Engineering, University of Houston, Houston, TX, 77204, USA (e-mail: mtuo@uh.edu; Xingpeng.Li@asu.edu).

even true for wind power plants taking advantage of kinetic energy stored in wind turbines.

With more generation coming from converter-based resources, insufficient inertia would be a main challenge for power systems stability. Moreover, low inertia feature together with variable renewable generation characteristics may lead to high rate of change of frequency (RoCoF) and large frequency excursion. When RoCoF or frequency deviation exceeds certain thresholds, protection devices would disconnect generators from the grid [3]. In fact, RoCoF related protection was found to be one of the main factors that limit the shift towards a 100% converter-based power system in Ireland [4]. If the frequency drops rapidly due to insufficient inertia, conventional generators may not be fast enough to respond to the deviation resulting in lower nadir and activation of load shedding. On August 9, 2019, over a million customers were affected by a major power disruption (mainly came from off-shore wind farms) that occurred across England and Wales and some parts of Scotland, and the frequency of the system hit 48.8 Hz [5]. The impact of RESs integration on power systems has been studied to address frequency stability challenge. In [6], the authors explored the massive deployment of grid-forming converters and its effects on frequency stability, results show that the system stability is lost when converter-based resources penetration reaches 80%. The Electric Reliability Council of Texas (ERCOT) studied the effect of low inertia on grid security and reliability [7]-[8]. To ensure frequency stability, the enhanced frequency response has been introduced in Great Britain recently which includes technologies like battery storage, interconnectors, and demand response [9]. Frequency control ancillary services have been implemented in the Australian National Electricity Market to maintain the system stability [10].

Following a different approach, several transmission system operators impose extra inertia constraints in the conventional unit commitment model to keep the minimum amount of synchronous inertia online [11]. Federal Energy Regulatory Commission has suggested that frequency control capabilities will need to be imposed into traditional unit commitment [12]. EirGrid has also introduced a synchronous inertial response (SIR) constraint to ensure that the available inertia does not fall below a static limit of 23,000 MWs in Ireland [13]. The optimal power flow with primary frequency response related constraint is investigated in [14]. Refs. [15] and [16] derived the analytical expression of the system frequency response model, frequency-related constraints are incorporated into traditional unit commitment: enforcing limitations on RoCoF, frequency nadir and steady-state error that are derived from a uniform frequency response model. A novel mixed-integer linear programming (MILP) unit commitment formulation was proposed in [17], which simultaneously optimizes energy production and the allocation of inertia. Despite these great efforts of modeling the classic system frequency response, previous studies rely on a simplification of the actual frequency dynamics and collective performance of system that cannot be able to capture the entire system characteristics.

The distinct frequency responses experienced by different bus have been observed in recent publications. System equivalent model-based operations may fail to supply sufficient ancillary services against contingency. Refs. [18] has also shown that generators on the buses adjacent to the event may suffer higher RoCoF comparing to distant buses. Similar conclusion has been made in [19] that the relative location of measurement point to disturbance is a pertinent factor in system inertial response, the RoCoF is usually higher for location where networks are weakly interconnected. To guarantee frequency stability accounting for spatial variation, a mixed analytical-numerical approach based on multi-regions has been studied in [20], which investigated the model combining evolution of the center of inertia and certain inter-area oscillations, stability conditions are proposed to co-optimize the existing ancillary services. Regarding oscillations mitigations, network coherency was considered as an alternative performance metric in [21], and a system dynamic model was formulated to determine the optimal allocation of additional inertia based on the disturbance location against high nodal frequency excursions. However, these approaches oversimplify the problem as they neglect the impact of disturbance propagation on locational inertial response. Furthermore, none of the aforementioned works discusses the criterion for worst contingency and the impact of event location. In addition, the sensitivity analysis of RES penetration level and the economic impact of virtual inertia have not been discussed thoroughly

In this article, we aim to bridge the aforementioned research gaps. Considering the literature review, the main contributions are summarized below:

- First, in contrast to existing literature such as [22] where only system equivalent model based RoCoF constrained SCUC (ERC-SCUC) are studied, we propose a novel location based RoCoF constrained SCUC (LRC-SCUC) model that can capture the locational frequency response characteristics and counteract the impact of system oscillation, guaranteeing the RoCoF security following a $G$-1 disturbance.

- Secondly, unlike [23] where initial RoCoF is assumed as highest value, a multiple-measurement-window method is introduced in this work to constrain highest value during oscillation. Besides, a piecewise linearization (PWL) based method is then proposed to convert the non-linear frequency constraints into linear frequency constraints in the proposed LRC-SCUC model, which allows us to optimally schedule the synchronous inertia as well as inertial services provided by non-synchronous resources to meet the minimum system inertia requirement for power systems with higher RES integration.

- Thirdly, we propose two virtual inertia-based LRC-SCUC (VI-LRC-SCUC) and ERC-SCUC (VI-ERC-SCUC) models to examine the effect of virtual inertia on inertia pricing and market efficient. Market results show that incorporating virtual inertia can largely reduce the system overall congestion and improve the market efficiency of the proposed LRC-SCUC model.

The remaining part of this paper is organized as follows. In Section II, we derive the post-contingency frequency dynamics and incorporate the corresponding analytic expressions into the SCUC model. Section III investigates the $G-1$ contingency at different locations to implement the non-linear location based RoCoF constraints, and the PWL method is utilized to linearize those non-linear location based RoCoF constraints.

Section IV presents the proposed LRC-SCUC model and VI-LRC-SCUC model, as well as the T-SCUC, ERC-SCUC and VI-ERC-SCUC as benchmark models. Case studies are presented in Section V. Finally, Section VI draws the main conclusions and describes potential future work.

## II. SYSTEM FREQUENCY DYNAMICS MODEL

The synchronous generator provides inertia to the power system through strongly coupled mechanical dynamics and electrical dynamics. Following a sudden change in load or a generation contingency, the dynamic behavior of the system frequency can be described using the swing equation of system equivalent single-machine representation,

$$P_m - P_e = M\frac{\partial \Delta \omega}{\partial t} + D\Delta \omega \tag{1}$$

where $M$ and $D$ are the aggregated system inertia constant and damping coefficient corresponding to the committed synchronous generators respectively. $P_m$ is the mechanical input power. $P_e$ is the electrical output power.

The transmission network can be modeled as a graph consisting of nodes (buses) and edges (branches). Using the topological information and the system parameters, the swing equation can then be extended and reformulated to describe the oscillatory behavior of each individual bus as follows [21],

$$m_i\ddot{\theta}_i + d_i\dot{\theta}_i = P_{in,i} - P_{e,i}, \ i \in \{1,...,n\} \tag{2}$$

where $m_i$ and $d_i$ denote the inertia coefficients and damping ratio for node $i$ respectively, while $P_{in,i}$ and $P_{e,i}$ refer to the power input and power output respectively. The electrical output power $P_{e,i}$ at bus $i$ is only related to the voltage phase angles as illustrated by (3).

$$P_{e,i} = \sum_{j=1}^{n} b_{ij} \sin (\theta_i - \theta_j), \ i \in \{1,...,n\} \tag{3}$$

By substituting (3) into (2), the network-coupled dynamical systems defined by sets of differential equations of the form,

$$m_i\ddot{\theta}_i + d_i\dot{\theta}_i = P_{in,i} - \sum_{j=1}^{n} b_{ij} \sin (\theta_i - \theta_j), \tag{4}$$
$$i \in \{1,...,n\}$$

With inertia on certain nodes $m_i > 0$, it is an approximation model for the swing dynamics of high-voltage transmission network within a few seconds after the event [24]. In this transient time interval, the network is justified as power system where the ratio factor of branch reactance to its resistance is high. Reactive power and voltage magnitude are not of concern, voltage amplitudes of the system are considered constant. The linear approximation $\sin (\theta_i - \theta_j) \approx \theta_i - \theta_j$ can be justified since the angle difference across the branch are small. Then eliminating passive load buses via Kron reduction [25], we can obtain a network-reduced model with $N$ generator buses. The phase angle $\theta$ of generator buses can be expressed by the following dynamic equation,

$$M\ddot{\theta} + D\dot{\theta} = P - L\theta \tag{5}$$

where $M = \text{diag}(\{m_i\})$, $D = \text{diag}(\{d_i\})$. Thus, for the Laplacian matrix $L$ of the grid, its off-diagonal elements are $l_{ij} = -b_{ij}V_i^{(0)}V_j^{(0)}$ and diagonals are $l_{ii} = \sum_{j=1, j\neq i}^{n} b_{ij}V_i^{(0)}V_j^{(0)}$. In this paper, the Laplacian matrix of the network-reduced model is real and symmetric [26], which has a complete orthogonal set of eigenvectors $\{\beta_\alpha\}$ with eigenvalue $\{\lambda_\alpha\}$. Due to the construction of the Laplacian matrix $L$ we have zero row and column sums, which implies that there is $\lambda_1 = 0$, corresponding to an eigenvector with constant elements, $(\beta_1)^T = \{\frac{1,...,1}{\sqrt{N_g}}\}$. Higher eigenvalues $\lambda_\alpha$ of $L$ are strictly positive for $\alpha = \{2, ..., N_g\}$, where $N$ is the number of generator buses, and the second-smallest eigenvalue is the algebraic connectivity [25]. Approach proposed in [24], [27] have shown high accuracy and robustness in identifying the critical nodes accounting for spatial inertia distribution. Under the assumption of homogeneous inertia, the frequency deviations at bus $i$ can then be derived,

$$\delta\dot{\theta}_i(t) = \frac{\Delta P e^{-\frac{\gamma t}{2}}}{m}\sum_{\alpha=1}^{N_g}\left(\beta_{\alpha i}\beta_{\alpha b}\frac{\sin\left(\sqrt{\frac{\lambda_\alpha}{m}-\frac{\gamma^2}{4}}\ t\right)}{\sqrt{\frac{\lambda_\alpha}{m}-\frac{\gamma^2}{4}}}\right) \tag{6}$$

where $m$ denotes average inertia distribution on generator buses; and bus $b$ is the location where disturbance occurs. With both damping and inertia being bounded at each bus, the ratio of damping coefficient to inertia coefficient is strictly prescribed within narrow ranges, $\gamma = d_i/m_i$, which is assumed as a constant [28]. As the frequency is monitored at discrete time interval, we consider a measurement window length of $\Delta t$, then RoCoF on bus $i$, $R_i(t)$, can be calculated as:

$$R_i(t) = -\frac{\delta\dot{\theta}_i(t+\Delta t) - \delta\dot{\theta}_i(t)}{2\pi\Delta t} \tag{7}$$

After substituting (6) into (7), we can obtain,

$$R_i(t) =$$
$$\frac{\Delta P e^{-\frac{\gamma t}{2}}}{2\pi m}\sum_{\alpha=1}^{N_g}\frac{\beta_{\alpha i}\beta_{\alpha b}}{\sqrt{\frac{\lambda_\alpha}{m}-\frac{\gamma^2}{4}}\Delta t}\left[\begin{array}{c}e^{-\frac{\gamma\Delta t}{2}}\sin\left(\sqrt{\frac{\lambda_\alpha}{m}-\frac{\gamma^2}{4}}(t+\Delta t)\right)\\-\sin\left(\sqrt{\frac{\lambda_\alpha}{m}-\frac{\gamma^2}{4}}t\right)\end{array}\right] \tag{8}$$

Expression (9) is the position-independent contribution $R_i^{(1)}$, which is the term corresponding to $\alpha = 1$ in (8). It can be observed that term $\alpha = 1$ describes the dynamics of system equivalent model.

$$R_i^{(1)}(t) = \frac{\Delta P e^{-\gamma t}(1 - e^{-\gamma\Delta t})}{2N_g\pi m\gamma\Delta t} \tag{9}$$

The initial inertial response can then be calculated as,

$$\lim_{\Delta t\to 0,\ t\to 0} R_i^{(1)}(t) = \frac{\Delta P}{2\pi N_g m} \tag{10}$$

which corresponds to the inertia response of system equivalent model. In (8)-(10), the inertia coefficient $m$ plays a vital role in the first few seconds which is inversely proportional to the RoCoF; higher inertia coefficient results in mitigated RoCoF value and oscillation amplitude, which allows longer responding time for the primary control to act.

Further study also suggests that amplitude and period of higher oscillations correspond to terms $\alpha > 1$ all depend on $\sqrt{\lambda_\alpha/m - \gamma^2/4}$. For IEEE 24-bus system, $\sqrt{\lambda_\alpha/m - \gamma^2/4} \in [0.88,\ 14.18]$, $\sqrt{\lambda_{\alpha \geq 3}/m - \gamma^2/4}$ is almost twice large than the second one. High-lying terms with larger eigenvalues $\lambda_\alpha$ contribute much less than the second slowest eigenmode, i.e. the Fielder mode, of the system Laplacian matrix $L$ [27]. Meanwhile higher-lying modes have short-period contributions, therefore only the first few eigenmodes of the network Laplacian, corresponding to its lower eigenvalues, mainly impact the value. In this study, we neglect them in the qualitative discussions and consider the RoCoF contributions from terms $\alpha \le 2$. The simplified but effective function to calculate nodal RoCoF is given as follows,

$$R_i^{(1,2)}(t) =$$

$$\frac{\Delta P e^{-\gamma t}(1-e^{-\gamma \Delta t})}{2N_g \pi m \gamma \Delta t} + \frac{\Delta P e^{-\frac{t}{2}}}{2\pi m} \frac{\beta_{2i}\beta_{2b}}{\sqrt{\frac{\lambda_2}{m}-\frac{\gamma^2}{4}}\Delta t}$$

$$\left[e^{-\frac{\Delta t}{2}}\sin\left(\sqrt{\frac{\lambda_2}{m}-\frac{\gamma^2}{4}}(t+\Delta t)\right)-\sin\left(\sqrt{\frac{\lambda_2}{m}-\frac{\gamma^2}{4}}t\right)\right]$$

(11)

By monitoring the average frequency change, the RoCoF relays can make more secure decisions during contingency. Practically, the time interval or measuring window for calculating RoCoF ranges from 5 cycles to 10 cycles [29]. In this paper, the average frequency change over a period of 100 ms (6 cycles) is defined as the RoCoF value.

## III. FORMULATION OF FREQUENCY CONSTRAINTS

We first focus on the impact of oscillations, RoCoF experienced by different generator buses can be different depending on their inertia coefficient and electrical distance from the disturbance, and the concept of local RoCoF is proposed in [30], generators in some areas would then suffer much higher RoCoF and have a higher chance to get tripped [31].

Fig. 1 shows the distribution of Fielder mode of the Laplacian matrix corresponding to the reduced network of the IEEE 24-bus system via Kron reduction. The original model has nodes where inertia $m_i = 0$ which gives the celebrated Kuramoto model on a network where angle differences become very large, which is not suitable for high-voltage electric power system analysis in this work [24]. Fielder mode of the reduced model manifests the frequency dynamics of individual generator bus regarding system connectivity and disturbance propagation [32].

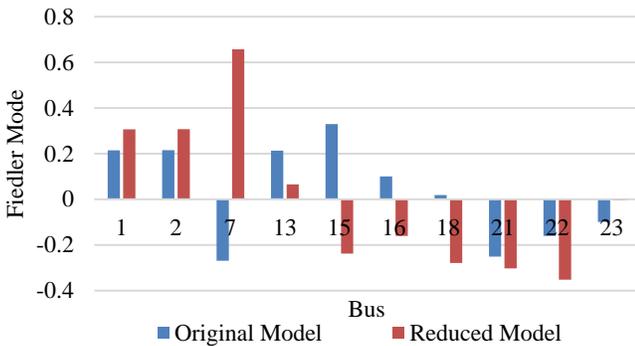

Fig. 1. Fielder mode distribution.

To investigate how Fielder mode affects the locational inertial response and the operation of relating ancillary services, simulations are conducted on the IEEE 24-bus system with a step increase of 180 MW load on bus 18. Dynamic simulation results are depicted in Fig. 2(a), as we can see that the frequency deviations at bus 13 and bus 23 are close to the deviation at center of inertia (COI) within the whole monitoring course. The largest deviation is observed at bus 7 corresponding to the highest Fielder mode absolute value.

These deviations in frequency can influence the locational frequency at different buses depending on the size and configuration of the system. Traditionally, generators adjacent to the event is assumed perceive highest RoCoF comparing to the generators on distant buses [6]. While results of [20] and [27] have emphasized the impact of oscillation and disturbance propagation, buses in distance may experience highest RoCoF after a time delay instead of at the initial time point. From the expression (10), it can be inferred that local buses sharing the same positive or negative Fielder mode with the event bus would perceive larger deviation before COI frequency, while the non-local generator buses with opposite positive and negative Fielder mode may experience highest RoCoF later than the COI frequency. Simulation results shown in Fig. 2(b) prove our inference. Following a contingency, there are larger frequency excursions and higher RoCoF on non-local generator buses to the disturbance comparing to COI frequency. For these non-local generator buses of opposite positive and negative Fielder mode value with disturbance bus, highest RoCoF occurs with a time period delay due to the disturbance propagation.

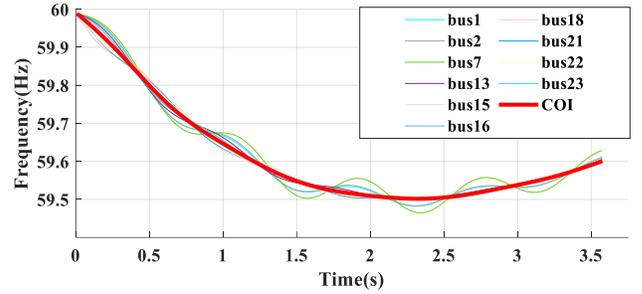

(a) Frequency for period between $t$=0 and $t$=3.5s.

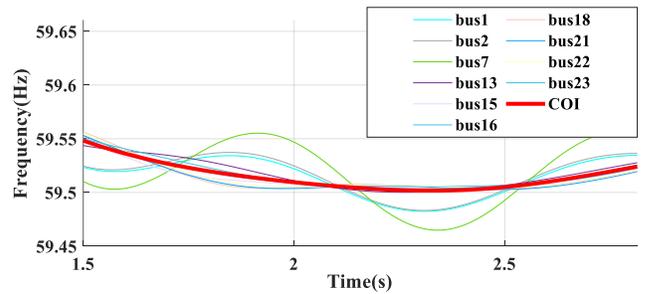

(b) Frequency for the period between $t$=1.5s and $t$=2.5s.

Fig. 2. Frequency response following a disturbance on bus 18.

To guarantee the locational frequency security for all buses and mitigate the highest RoCoF during the oscillation, a multiple-measurement-window method is introduced to capture the highest RoCoF during the oscillation. Compared to the case of a sudden load increase, the loss of generation not only causes mismatch in system power balance, but also the degradation of the system synchronous inertia, resulting in even higher frequency deviation and larger initial RoCoF. Thus, the $G$-1 contingency of largest generation is considered as the worst contingency in this paper. And the aforementioned

RoCoF expression in (11) are then incorporated as constraints in SCUC, the resulting locational RoCoF constraint (12) that respects the prescribed threshold $RoCoF_{lim}$ is then reformulated as follows with $t$ being set as $T_1$ and $T_2$. These two typical setting points are same for all generator buses, which is mainly based on oscillation period of second lowest term $\alpha = 2$, considering trade-off between computational efficiency and measurement accuracy [20].

$$\frac{\Delta P e^{-\frac{t}{2}}(1 - e^{-\gamma\Delta t})}{2N_g \pi m_\Delta \gamma \Delta t} + \frac{\Delta P e^{-\frac{t}{2}}}{2\pi m_\Delta} \frac{\beta_{2i}\beta_{2b}}{\sqrt{\frac{\lambda_2}{m_\Delta} - \frac{\gamma^2}{4}}\Delta t}$$

$$\left[ \begin{array}{c} e^{-\gamma\frac{\Delta t}{2}}\left(\sqrt{\frac{\lambda_2}{m_\Delta} - \frac{\gamma^2}{4}}\left(t + \Delta t\right)\right) \\ -\sin\left(\sqrt{\frac{\lambda_2}{m_\Delta} - \frac{\gamma^2}{4}}t\right) \end{array} \right] \geq RoCoF_{lim} \quad (12)$$

Where $m_\Delta = m - \Delta m$, $\Delta m$ denotes an average loss of inertia coefficient distributed on each bus, $N_g$ is the number of generator buses. Within a plausible range of variables $(\Delta P, m, \Delta m, \gamma)$, the RoCoF on all generator buses following a $G - 1$ contingency can be calculated with the left-hand side terms of (12).

## IV. SCUC FORMULATIONS

The objective of SCUC is to minimize operational cost of generators subject to various constraints. Traditional SCUC (T-SCUC) model does not consider frequency constraints, which may lead to severe system instability for lower-inertia power grid with substantial amount of renewable generation. In this section, we first introduce the T-SCUC and then describe how to incorporate RoCoF related constraints into the ERC-SCUC and the proposed LRC-SCUC. Subsequently, we will describe the proposed VI-LRC-SCUC model that utilize virtual inertia.

Objective function (13a) is shared by three SCUC models: T-SCUC, ERC-SCUC and LRC-SCUC. It is to minimize the total system cost consisting of variable fuel costs, no-load costs, start-up costs, and reserve costs.

The T-SCUC model includes various constraints (13b)-(13o). Equation (13b) enforces the nodal power balance. Network power flows are calculated in (13c) and are restricted by the transmission capacity as shown in (13d). The scheduled energy production and generation reserves are bounded by unit generation capacity and ramping rate (13e)-(13j). As defined in (13h), the reserve requirements ensure the reserve is sufficient to cover any loss of a single generator. The start-up status and on/off status of conventional units are defined as binary variables (13k)-(13o).

$$\min_\Phi \sum_{g \in G}\sum_{t \in T}\left(c_g P_{g,t} + c_g^{NL} u_{g,t} + c_g^{SU} v_{g,t} + c_g^{RE} r_{g,t}\right) \quad (13a)$$

$$\sum_{g \in G} P_{g,t} + \sum_{k \in K(n-)} P_{g,t} - \sum_{k \in K(n+)} P_{g,t} - D_{n,t} \\ + E_{n,t} = 0, \ \forall n, t, \quad (13b)$$

$$P_{k,t} - b_k(\theta_{n,t} - \theta_{m,t}) = 0, \ \forall k, t, \quad (13c)$$

$$-P_k^{\max} \leq P_{k,t} \leq P_k^{\max}, \ \forall k, t, \quad (13d)$$

$$P_g^{\min} u_{g,t} \leqslant P_{g,t}, \ \forall g, t, \quad (13e)$$

$$P_{g,t} + r_{g,t} \leq u_{g,t} P_g^{\max}, \ \forall g, t, \quad (13f)$$

$$0 \leq r_{g,t} \leq R_g^{re} u_{g,t}, \ \forall g, t, \quad (13g)$$

$$\sum_{j \in G} r_{j,t} \geq P_{g,t} + r_{g,t}, \ \forall g, t, \quad (13h)$$

$$P_{g,t} - P_{g,t-1} \leq R_g^{hr}, \ \forall g, t, \quad (13i)$$

$$P_{g,t-1} - P_{g,t} \leq R_g^{hr}, \ \forall g, t, \quad (13j)$$

$$v_{g,t} \geq u_{g,t} - u_{g,t-1}, \quad \forall g, t, \quad (13k)$$

$$v_{g,t+1} \leq 1 - u_{g,t} \quad \forall g, t \leq nT - 1, \quad (13l)$$

$$v_{g,t} \leq u_{g,t} \quad \forall g, t, \quad (13m)$$

$$v_{g,t} \in \{0, 1\}, \ \forall g, t, \quad (13n)$$

$$u_{g,t} \in \{0, 1\}, \ \forall g, t, \quad (13o)$$

As mentioned before, the ERC-SCUC model is constrained to guarantee generator frequency stability considering the relative location to the potential $G - 1$ contingency. Loss of synchronous generation would lead to a reduction in total system inertia which would cause a higher RoCoF comparing to the event with no synchronous inertia loss. The system RoCoF limit following a $G - 1$ contingency is guaranteed by applying the following set of frequency related constraints (13p)-(13r). (13p) defines the rated power of dispatched generators, while (13q) calculates the system synchronous inertia. (13r) is derived from the swing equation of system equivalent model, the constraint ensures the system frequency security against loss of generation.

$$k_{g,t} = P_g^{\max} u_{g,t}, \qquad \forall g, t, \quad (13p)$$

$$M_t = \frac{\sum_{g \in G} 2H_g k_{g,t}}{\omega_0}, \qquad \forall t, \quad (13q)$$

$$-P_{g,t}/(M_t\omega_0 - 2H_g k_{g,t}) \geq RoCoF_{lim} \quad \forall g, t, \quad (13r)$$

If the generation scheduling is not appropriately constrained to reflect the $G - 1$ contingency and distinct locational frequency, unexpected tripping of RoCoF relays and cascading contingency may take place. To address this issue, squared Fiedler mode amplitude $\beta_{2b}^2$ is introduced in this paper to assess the impact of disturbance, a power loss at bus $b$, on frequencies in the whole grid. Considering the inter oscillation between area which may cause unexpected tripping of RoCoF relays, we create two locational RoCoF constraints in this paper based on the definition of local buses and non-local buses. Constraint (13s)-(13u) defines the average nodal inertia change over generator buses due to loss of a generator. Constraints (13v) and (13w) ensure system stability by imposing limit on locational highest captured RoCoF of buses under possible worst $G - 1$ contingency.

$$\Delta m_{g,t} = \frac{2H_g k_{g,t}}{N_g \omega_0}, \qquad \forall g, t, \quad (13s)$$

$$m_t = \frac{M_t}{N_g}, \quad \forall t, \qquad (13t)$$

$$m_{g,t} = m_t - \Delta m_{g,t}, \qquad \forall g, t, \qquad (13u)$$

$$\frac{P_{g,t}e^{-\gamma T_1}(1-e^{-\gamma\Delta t})}{2N_g\pi m_{g,t}\gamma\Delta t} + \frac{P_{g,t}e^{-\gamma\frac{T_1}{2}}}{2\pi m_{g,t}}\frac{\beta_{2n}\beta_{2b}}{\sqrt{\frac{\lambda_2}{m_{g,t}} - \frac{\gamma^2}{4}}\Delta t}$$

$$\left[e^{-\gamma\frac{\Delta t}{2}}\left(\sqrt{\frac{\lambda_2}{m_{g,t}} - \frac{\gamma^2}{4}}(T_1+\Delta t)\right)\right.$$
$$\left. - \sin\left(\sqrt{\frac{\lambda_2}{m_{g,t}} - \frac{\gamma^2}{4}}T_1\right)\right] \leq -\text{RoCoF}_{\text{lim}}, \qquad (13v)$$

$$\forall n \in N_{loc}, g, t,$$

$$\frac{P_{g,t}e^{-\gamma T_2}(1-e^{-\gamma\Delta t})}{2N_g\pi m_{g,t}\gamma\Delta t} + \frac{P_{g,t}e^{-\gamma\frac{T_2}{2}}}{2\pi m_{g,t}}\frac{\beta_{2n}\beta_{2b}}{\sqrt{\frac{\lambda_2}{m_{g,t}} - \frac{\gamma^2}{4}}\Delta t}$$

$$\left[e^{-\gamma\frac{\Delta t}{2}}\left(\sqrt{\frac{\lambda_2}{m_{g,t}} - \frac{\gamma^2}{4}}(T_2+\Delta t)\right)\right.$$
$$\left. - \sin\left(\sqrt{\frac{\lambda_2}{m_{g,t}} - \frac{\gamma^2}{4}}T_2\right)\right] \leq -\text{RoCoF}_{\text{lim}}, \qquad (13w)$$

$$\forall n \in N_{n-loc}, g, t,$$

There are different implementations for synchronous machine response emulation. The study in [33] shows that control schemes can be utilized to provide equivalent inertia through non-synchronous devices. Such concept introduces techniques like virtual inertia emulation to mimic the behavior of synchronous machines, displacing the inertia provided by synchronous generators with cheap virtual ancillary service provided by various resources such as virtual synchronous machine. To study the effect of virtual inertia on the power grid, an aggregated virtual inertia involved location based RoCoF constrained SCUC model or a VI-LRC-SCUC model is proposed and examined in this paper. Compared to (13a), this model also considers the cost for virtual inertia provision; the updated objective function for the proposed VI-LRC-SCUC model is shown in (14a). Moreover, when virtual inertia is considered, constraints (13q)-(13r) should be replaced by (14b)-(14c). Expression (14b) describes the system inertial response respect to aggregate inertia contributions from condensers and inverter-based resources. (14c) defines the change in average nodal inertia while virtual inertia is applied.

$$\min_{\Phi} \sum_{g \in G}\sum_{t \in T}(c_g P_{g,t} + c_g^{NL}u_{g,t} + c_g^{SU}v_{g,t} + c_g^{RE}r_{g,t})$$
$$+ \sum_{t \in T}c^{VI}M_t^{VI} \qquad (14a)$$

$$-P_{g,t}/(M_t\omega_0 + M_t^{VI}\omega_0 - 2H_g k_{g,t}) \geq \text{RoCoF}_{\text{lim}},$$
$$\forall g, t, \qquad (14b)$$

$$m_t = \frac{M_t}{N_g} + \frac{M_t^{VI}}{N_g}, \quad \forall t, \qquad (14c)$$

The virtual inertia requires fast responsive energy buffer; the kinetic energy in a wind turbine and the energy in a battery are limited energy resources for virtual inertia responses. Hence, it is practical to set a limit on the total virtual inertia due to budget limit and resource limit. This limit should also be considered as follows,

$$0 \leq M_t^{VI} \leq M_t^{Total}, \quad \forall t, \qquad (14d)$$

This paper examines five different SCUC models that are summarized in TABLE I. In this table, the objective functions and constraints enforced are listed for each SCUC model.

TABLE I
SCUC FORMULATION OF DIFFERENT MODELS

| Model | Objective Function | Shared Constraints | Unique Constraints |
|---|---|---|---|
| T-SCUC | (13a) | (13b)-(13o) | None |
| ERC-SCUC | | | (13p)-(13r) |
| LRC-SCUC | | | (13p)-(13q), (13s)-(13w) |
| VI-ERC-SCUC | (14a) | | (13p)-(13q), (14b), (14d) |
| VI-LRC-SCUC | | | (13p)-(13q), (13s), (13t)-(13w), (14c)-(14d) |

Note that non-linear constraints (13v) and 13(w) are linearized with the PWL method described in Section III before solving the associated SCUC models.

The constraints on RoCoF for locational frequency dynamics are nonlinear. In order to incorporate these frequency-related constraints into the proposed LRC-SCUC model, a linear approximation method is introduced. [34] proposes a piece-wise linearization technique for obtaining a linearized expression. Since respective damping and droop gains are usually strictly prescribed proportional to the synchronous inertia, the RoCoF expression becomes a function of three variables $R_i^{(1,2)}(\Delta P, m, \Delta m)$. The expression has local convexity within the interval of interest, that can be expressed as $a_v^i\Delta P^\eta + b_v^i m^\eta + c_v^i\Delta m^\eta + d_v^i$. In order to determine these four coefficients, A heuristic least-squares method is proposed in [35] to solve this problem.

The least squares based PWL method aims to minimize the following objective function,

$$\min_{\Psi} \sum_{\eta}\Big(\max_{1 \leq v \leq \bar{v}}\{a_v^i\Delta P^\eta + b_v^i m^\eta + c_v^i\Delta m^\eta + d_v^i\}$$
$$- R_i^{(1,2)}(\Delta P^\eta, m^\eta, \Delta m^\eta)\Big)^2 \qquad (15)$$

where $\Psi = \{a_v, b_v, c_v, d_v, \forall v\}$ is the set of parameters to be determined; $i$ denotes the measurement bus; and $\eta$ denotes the evaluation point; $v$ refers to the index of PWL segments and $\bar{v}$ denotes the number of PWL segments. The problem of fitting $\max_{1 \leq v \leq \bar{v}}\{a_v^i\Delta P^\eta + b_v^i m^\eta + c_v^i\Delta m^\eta + d_v^i\}$ to $-R_i^{(1,2)}(\Delta P^\eta, m^\eta, \Delta m^\eta)$ over the plausible range can be considered as minimizing difference between the appropriate PWL segment and the RoCoF function. To solve this min-max problem and eliminate the inner max operator from the objective function, new variables $t_v$ are introduced and defined as follows,

$$t_1^i = \max\{a_1^i\Delta P^\eta + b_1^i m^\eta + c_1^i\Delta m^\eta + d_1^i,$$
$$a_2^i\Delta P^\eta + b_2^i m^\eta + c_2^i\Delta m^\eta + d_2^i\}, \qquad (16a)$$

$$t_{v-1}^i = \max\{t_{v-2}^i, a_v^i\Delta P^\eta + b_v^i m^\eta + c_v^i\Delta m^\eta$$
$$+ d_v^i\}, 3 \leq v \leq \overline{v}. \quad (16b)$$

Adding new linear inequalities would relieve the objective function from the "max" operator [34]. We basically introduce $\overline{v} - 1$ new binary $\alpha_v$ as well as $\overline{v} - 1$ continuous variables $t_v^i$, the unconstrained min-max optimization problem (15) can be reformulated as a constrained optimization problem (17)-(18), which is the RoCoF linearization problem.

$$\min_\Psi \sum_\eta (t_{\overline{v}-1}^i (\Delta P^\eta, m^\eta, \Delta m^\eta) - R_i^{(1,2)}(\Delta P^\eta, m^\eta,$$
$$\Delta m^\eta))^2 \quad (17)$$

subject to the following constraints,

$$a_1^i\Delta P^\eta + b_1^i m^\eta + c_1^i\Delta m^\eta + d_1^i \leq t_1^i \leq a_1^i\Delta P^\eta +$$
$$b_1^i m^\eta + c_1^i\Delta m^\eta + d_1^i + \alpha_1^i\Omega, \quad \forall\eta, \quad (18a)$$

$$a_2^i\Delta P^\eta + b_2^i m^\eta + c_2^i\Delta m^\eta + d_2^i \leq t_1^i \leq a_2^i\Delta P^\eta +$$
$$b_2^i m^\eta + c_2^i\Delta m^\eta + d_2^i + (1 - \alpha_1^i)\Omega, \quad \forall\eta, \quad (18b)$$

$$t_{v-2}^i \leq t_{v-1}^i \leq t_{v-2}^i + \alpha_{v-1}^i\Omega, \quad \forall\eta, 3 \leq v \leq \overline{v}, \quad (18c)$$

$$a_v^i\Delta P^\eta + b_v^i m^\eta + c_v^i\Delta m^\eta + d_v^i \leq t_{v-1}^i \leq a_v^i\Delta P^\eta +$$
$$b_v^i m^\eta + c_v^i\Delta m^\eta + d_v^i + (1 - \alpha_{v-1}^i)\Omega, \quad (18d)$$
$$\forall\eta, 3 \leq v \leq \overline{v}.$$

where $\Omega$ is a sufficiently large positive number.

Upon obtaining the optimal solution $(a_v^*, b_v^*, c_v^*, d_v^*)$ of the nodal RoCoF linearization model, the nonlinear location based RoCoF constraints (13v) and (13w) can be converted into linear constraints in SCUC: $4*(\overline{v}-1)$ inequalities (18) that ensures $t_{\overline{v}-1}^i$ is maximum, along with the RoCoF threshold constraints $t_{\overline{v}-1}^i \leq -\text{RoCoF}_{\text{lim}}$.

## V. CASE STUDIES

### A. Test System Description

A case study on IEEE 24-bus system is provided to demonstrate the effectiveness of the proposed methods [36]. This test system contains 24 buses, 38 generators and 38 lines, which also considers decarbonized generation characterized by wind power and solar power on generator buses. Fig. 3 shows the renewable generation and load profile for a system scenario with 60% of maximum renewable energy penetration level during peak hour. Electricity demand ranges from 1,432 MW to a peak of 3,222 MW. The mathematical models are implemented in Python using Pyomo [37]-[38] and solved with the Gurobi solver [39], and the optimality gap is set to 0.1%. The computer with Intel® Xeon(R) W-2195 CPU @ 2.30GHz and 128 GB of RAM was utilized to conduct the numerical simulations.

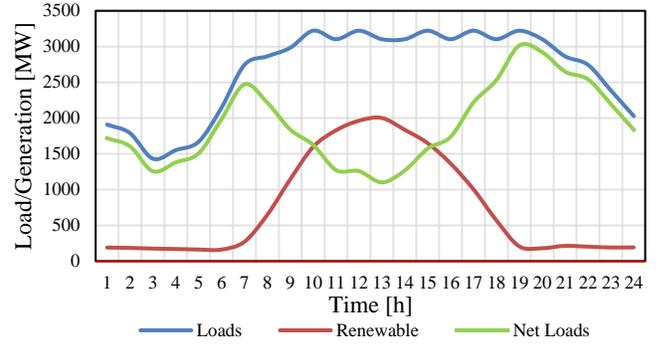

Fig. 3. Renewable generation and load profile of the IEEE 24-bus system.

### B. Investigation of Frequency Propagation

In this subsection, the effect of Fiedler mode on disturbance propagation and locational inertial response is simulated and studied. We first investigate the implications of a generation loss in scenarios with different system inertia. The locational RoCoF is numerically calculated based on expression (11), RoCoF at $t = 0$s and $\Delta t = 0.1$s on local bus 21 adjacent to event bus 18 that is illustrated in Fig. 4 where the loss of largest unit is considered. It can be inferred that the initial frequency oscillations are affected by the location of disturbance.

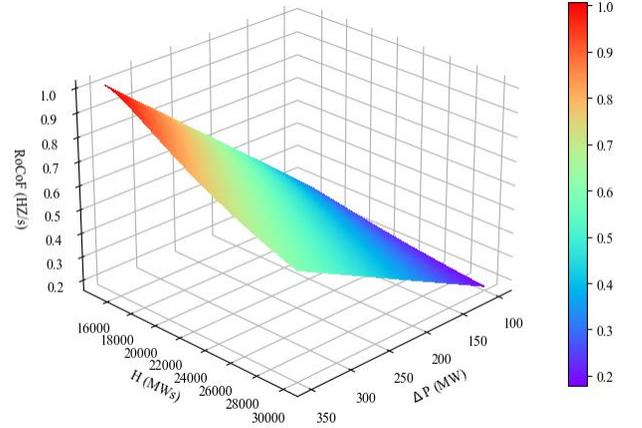

Fig. 4. RoCoF of bus 21 at $t = 0$s and $\Delta t = 0.1$s following a $G$-1 contingency on bus 18.

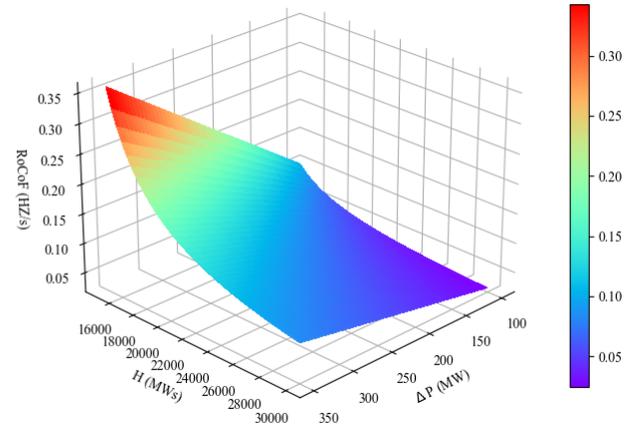

Fig. 5. RoCoF of bus 1 at $t = 0$s and $\Delta t = 0.1$s following $G$-1 contingency on bus 18.

The inertial response on non-local bus 1 is depicted in Fig. 5, generator buses with opposite positive and negative Fiedler mode experience a much smaller initial RoCoF at $t = 0$s.

 

Combining results Fig. 5 and Fig. 6, it can be observed that local buses adjacent to disturbance substantially experience much higher initial RoCoF than other buses under at $t = 0s$. Hence, the generators buses adjacent to disturbance with large Fiedler mode are more likely to violate the RoCoF limits at the initial time point following a $G - 1$ contingency.

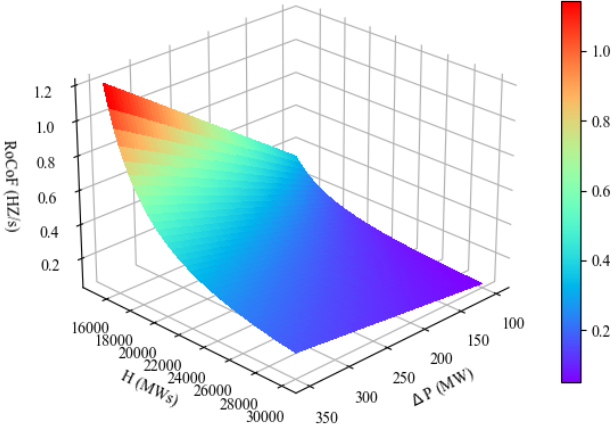

Fig. 6. RoCoF of bus 1 at $t = 0.4s$ and $\Delta t = 0.1s$ following $G - 1$ contingency on bus 18.

The numerically calculated RoCoF for bus 1 at $t = 0.4s$ and $\Delta t = 0.1s$ under the same scenario is plotted in Fig. 6. It is interesting to find that local bus 1 experiences much large RoCoF at $t = 0.4s$ than same RoCoF at $t = 0.0s$. The results show that only considering the initial RoCoF may fail to capture highest locational RoCoF value during the oscillation, simulation results indicate the effectiveness of the proposed multiple-measurement-window method which allows us to incorporate the captured the highest RoCoF into LRC-SCUC model and subsequently secure locational frequency stability.

It is also important to note that when system inertia drops below a certain threshold the RoCoF increases drastically especially for the buses have relatively large Fiedler mode absolute value. This subsection demonstrates the effect of inertia and Fiedler mode value on locational RoCoF: with higher system inertia, the amplitude of the oscillation decreases especially for buses with large Fiedler mode absolute value.

### C. LRC-SCUC Model

We set the system nominal frequency to 60 Hz. Regarding post-contingency frequency limits, RoCoF must be higher than -0.5Hz/s to avoid the tripping of RoCoF-sensitive protection relays. In [23], generator settings have a typical damping to inertia ratio ranges from [0.086, 0.133]. Similar dynamic parameters are utilized in this work, the ratio is subsequently considered constant as 0.1 in this work. We first conduct the simulation of T-SCUC model for 24-hour period, which serves as a benchmark to show the impact of RoCoF constraints. The evaluation point and the PWL segments are set with $\eta = 3$ and $\bar{v} = 4$ for accuracy and computational efficiency [23]. The computational time for the T-SCUC model is 95.61s while it decreases to 20.91s for ERC-SCUC model. A possible reason is that frequency related constraints reduce the size of feasible solution set, infeasible solutions are discarded earlier because of violating the RoCoF constraints. Meanwhile, most RoCoF related constraints in LRC-SCUC model are not binding constraints. With PWL formulation solved offline, the proposed LRC-SCUC model has a computational time of 75.45s, indi-cating that the proposed LRC-SCUC model can be solved efficiently.

Fig. 7 presents the evolutions of aggregated system inertia over the scheduling horizon. When net load increases during the period of hours 5-8 and hours 16-20, the total system aggregated inertia increases as well for all three cases. Compared to T-SCUC model, imposing RoCoF constraints leads to more synchronous generators scheduled online to ensure minimum synchronous inertia online. The inertia of committed synchronous generators for T-SCUC model is much lower than ERC-SCUC and LRC-SCUC models where RoCoF related constraints are implemented. Note that the total system synchronous inertia of the proposed LRC-SCUC model is the highest among all three models, which reflects the impact of imposing locational RoCoF constraints.

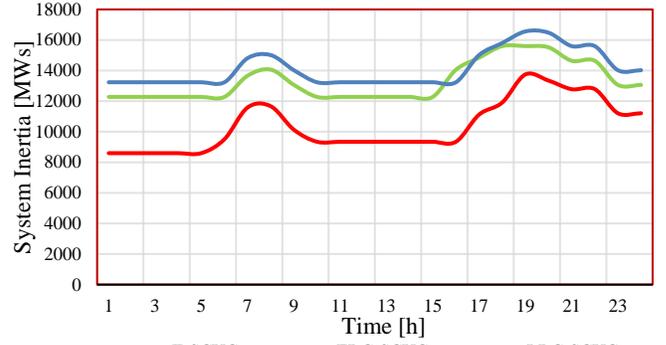

Fig. 7. Impact of RoCoF constraints on the total system inertia.

Some other insightful conclusions can also be drawn from Fig. 7. When net load decreases dramatically between hours 7-13 due to the increase of RES generation, the system aggregate inertia in T-SCUC case drops significantly subsequently, contingency happens in this time interval may lead to larger frequency deviation and higher RoCoF. While system inertia maintains relatively high level in two other cases indicating mitigated RoCoF and frequency deviations.

Furthermore, we investigate how commitment scheduling impacts the system frequency dynamics following $G - 1$ contingency [40]. The worst-case contingency is assumed to take place in peak hour 12, time-domain simulations are then conducted for all three models on Transient Security Analysis Tools (TSAT) following the loss of largest generator. The detailed model is utilized in this work, and dynamic values are selected within appropriate ranges.

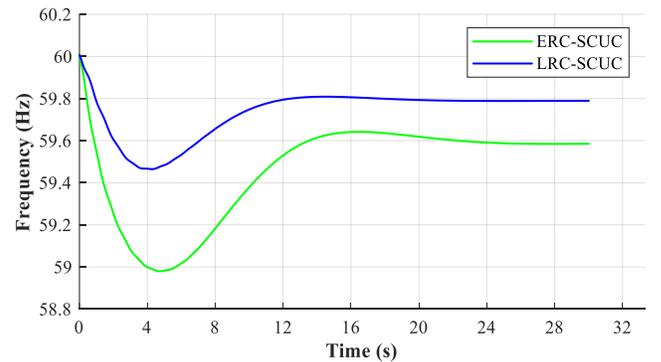

Fig. 8. System frequency response after loss of the generator with the largest generation at hour 12.

Results are plotted in Fig. 8 comparing the frequency dynamics for ERC-SCUCU and LRC-SCUC. In T-SCUC model without any RoCoF related constraints, loss of the largest committed generation of 400 MW on bus 22 leads to an extremely large frequency excursion at 54.1 Hz. For ERC-SCUC, loss of largest generation at 182.9 MW on bus 22 result in a frequency nadir at 58.9 Hz which is below the threshold of 59.3 Hz. The potential largest $G-1$ contingency level for LRC-SCUC is reduced to 155 MW, which is about 5% of largest demand. The system frequency is maintained above the threshold. Results indicate that the proposed LRC-SCUC model would help following primary frequency response, thus subsequently protect against the worst-case contingency risk.

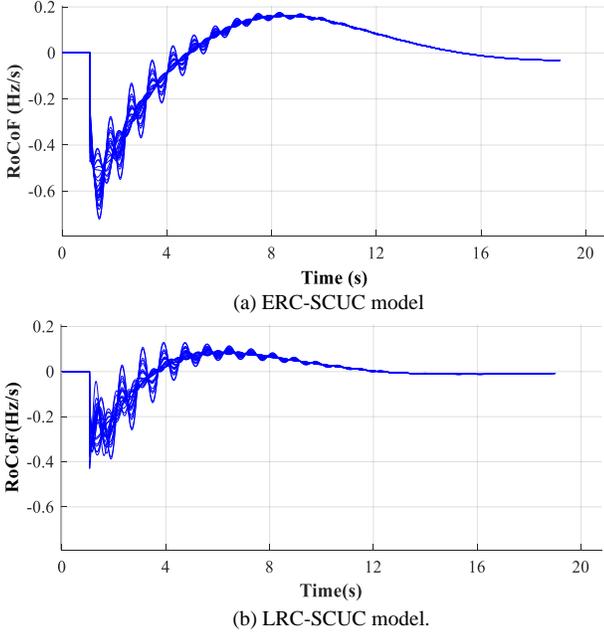

Fig. 9. RoCoF of all buses following the loss of largest generation in different cases.

(a) ERC-SCUC model

(b) LRC-SCUC model.

Fig. 9 compares the average RoCoF over a period of 100 ms on generator buses for ERC-SCUC and LRC-SCUC models following the worst-case contingency. For the T-SCUC results, the RoCoF on all buses violate the prescribed limit easily by a large margin especially in time interval from 9h to 16h. For ERC-SCUC model, the highest RoCoF derived from COI frequency can be numerically calculated as -0.50 Hz/s which is above the required value. However, the highest locational RoCoF still violates the prescribed RoCoF limit which reflects that impact of oscillations cannot be handled well by ERC-SCUC model. For the generator commitment and dispatch solution obtained from the proposed LRC-SCUC model, the lowest locational RoCoF is maintained above -0.5 Hz/s, meeting the RoCoF security requirement. It also can be observed that initial RoCoF may not be the lowest RoCoF within the oscillation period, imposing multiple measurement windows can help us capture highest RoCoF. The results indicate the effectiveness and necessity to incorporate locational RoCoF constraints into the scheduling model.

The results of three models are summarized in TABLE II. Imposing system equivalent model based RoCoF constraints leads to a 10.90% increase in total system cost, the total generation cost increases from $891,391 to $988,524. In the presence of location based RoCoF constraints, an increase of 17.90% in total operational cost is observed. On the other hand, the reserve cost decreases significantly as the worst-case contingency level mitigated. Meanwhile, additional synchronous machines are committed to cover the shortage of inertia to limit locational RoCoF following the loss of largest generation, which accordingly increases the operation cost as well as the start-up cost.

TABLE II
SCUC COSTS [$] UNDER DIFFERENT MODELS

| Model | Total | Start-up | Operation | Reserves |
|---|---|---|---|---|
| T-SCUC | 891,391 | 56,704 | 744,973 | 89,714 |
| ERC-SCUC | 988,524 | 59,462 | 881,633 | 47,429 |
| LRC-SCUC | 1,050,989 | 66,810 | 955,983 | 28,196 |

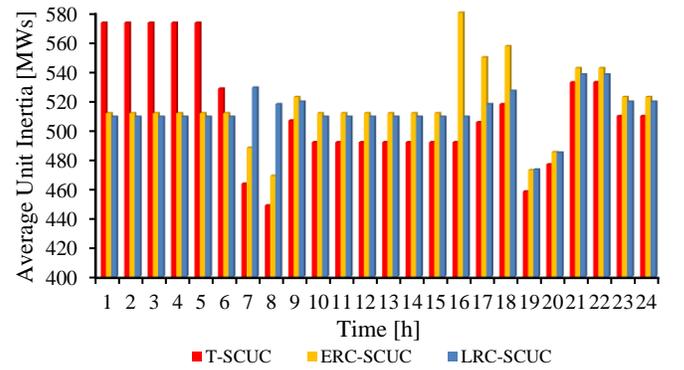

Fig. 10. Impact of RoCoF constraints on the average unit inertia contribution.

Furthermore, average unit inertia (AUI) of committed generators is compared in Fig. 10 over the whole dispatching horizon. A noticeable large value of AUI between hours 1-5 and hour 16-18 are monitored in T-SCUC and ERC-SCUC cases. Larger AUI indicates that generators of larger rated power and synchronous inertia are scheduled in the system operation, implying potential larger worst-case contingency level. As shown, the AUI of proposed LRC-SCUC case is relatively stable during the whole horizon. To achieve the energy system with more RES penetration level, re-dispatching existing generators provides options that can help mitigate the frequency deviation and RoCoF violations.

### D. Sensitivity Analysis on RES Penetration Levels

The sensitivity analysis with different renewable penetration levels is conducted in this subsection. Following the daily load profile shown in Fig. 3, four scenarios are considered for different RES penetration levels during peak hour from 20% to 80% with an increment of 20%. Given the system peak load is 3,222 MW at hour 12, the corresponding RES generations during peak hour are 644.4 MW, 1,288.8 MW, 1,933.2 MW and 2,577.6 MW respectively.

TABLE III
SYSTEM INERTIA [MWs] UNDER DIFFERENT MODELS

| RES Penetration Level | 20% | 40% | 60% | 80% |
|---|---|---|---|---|
| T-SCUC | 11,706 | 10,434 | 9,342 | 7,792 |
| ERC-SCUC | 14,639 | 13,063 | 12,275 | 10,880 |
| LRC-SCUC | 15,599 | 14,503 | 13,235 | 13,095 |

TABLE III presents the aggregated system inertia value at peak hour 12 in different scenarios. As RES penetration level

 

increases from 20% to 80%, the synchronous inertia of committed synchronous generators based on T-SCUC drops significantly comparing to the LRC-SCUC model.

| RES Penetration Level | 20% | 40% | 60% | 80% |
|---|---|---|---|---|
| T-SCUC | -1.26 | -1.32 | -1.45 | -1.65 |
| ERC-SCUC | -0.75 | -0.66 | -0.63 | -0.62 |
| LRC-SCUC | -0.34 | -0.38 | -0.42 | -0.46 |

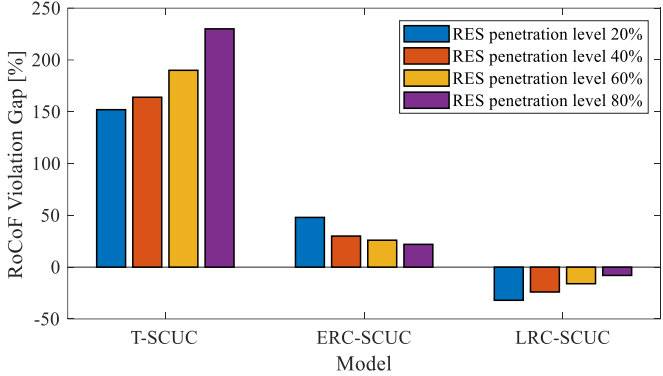

Fig. 11. RoCoF violation gaps for different scenarios.

One may be interested to compare the nodal frequency performance of all three SCUC models. Simulations have been conducted in TSAT considering worst-case contingency happens in peak hour 12. TABLE IV shows the highest RoCoF monitored on generator buses following the $G-1$ contingency under different scenarios. It can be observed that highest RoCoF in both ERC-SCUC and T-SCUC cases violates the limit which may subsequently trip frequency-sensitive protection relays, while the RoCoF is secured above predetermined threshold in all cases with the proposed LRC-SCUC model.

The RoCoF violation gap which indicates difference between the actual values and the thresholds are depicted in Fig. 11[23]. The violation gaps are all positive in T-SCUC and REC-SCUC indicating threshold violations. As RES penetration level increases, less conservative solutions are observed for LRC-SCUC model in this sensitivity test. Understandably, non-uniform distribution caused by generator aggregation is relieved because of less generator on.

| RES Penetration Level | 20% | 40% | 60% | 80% |
|---|---|---|---|---|
| T-SCUC | 1,357,203 | 1,077,439 | 891,391 | 750,199 |
| ERC-SCUC | 1,459,983 | 1,181,493 | 988,524 | 845,606 |
| LRC-SCUC | 1,546,833 | 1,253,221 | 1,050,989 | 908,353 |

The costs with those three SCUC models are listed in TABLE V. Obviously, incorporating frequency related RoCoF constraints in SCUC leads to an increase of system total generation cost in all scenarios. The change in the cost is dependent on the change in penetration level of renewable resources.

### E. Influence of Virtual Inertia

In this subsection, the proposed VI-LRC-SCUC model that considers virtual inertia $M_t^{VI}$ is implemented and the effects of virtual inertia on day-ahead scheduling and system stability are examined. We first consider $M_t^{Total} = \infty$ with no upper limit, the cost curves of VI-ERC-SCUC and VI-LRC-SCUC are depicted in Fig. 12. Results show that introducing virtual inertia would potentially reduce the total operational cost of VI-LRC-SCUC model, while cheaper inertia price leads to higher reduction value. With inertia price at 0$/MWs, the total operational cost of VI-LRC-SCUC is equal to the cost of T-SCUC, while it is close to the cost of LRC-SCUC when inertia price is over 0.75$/MWs.

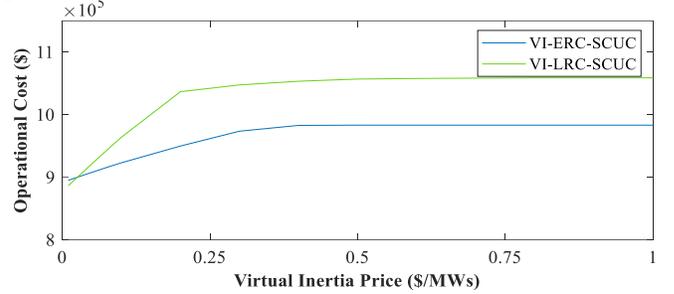

Fig. 12. Operational cost curve.

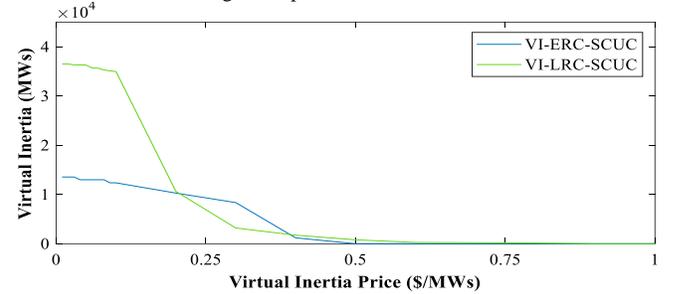

Fig. 13. Cost curve of virtual inertia provision.

The cost curve of virtual inertia provision has been depicted in Fig. 13. As can be seen, VI-LRC-SCUC is more sensitive to price change of virtual inertia than VI-LRC-SCUC model. The results also indicate that cheaper inertia price doesn't increase the value of needed virtual inertia in both cases. When virtual inertia price increase above 0.5$/MWs, the virtual inertia purchased by the system would decrease to zero, that means system would rather turn on conventional synchronous generators than purchasing additional virtual inertia.

Accounting for the available capacity and the total cost of the inertia-emulating devices, there is a total budget constraint for additional virtual inertia [21], the total available virtual inertia $M_t^{Total}$ is considered as 2,000 MWs. To evaluate the influence of including virtual inertia based RoCoF constraints in SCUC models, we analyze the average market results of five models over 24 hours.

| Model | Average LMP [$/MWh] | Average congestion LMP [$/MWh] | Energy LMP [$/MWh] |
|---|---|---|---|
| T-SCUC | 37.56 | 23.31 | 14.25 |
| ERC-SCUC | 41.44 | 24.20 | 17.24 |
| VI-ERC-SCUC | 40.41 | 23.17 | 17.24 |
| LRC-SCUC | 45.89 | 29.09 | 16.80 |
| VI-LRC-SCUC | 37.88 | 21.08 | 16.80 |

The average locational market price (LMP) is defined in [41], including energy component and congestion component is presented in TABLE VI. With virtual inertia implemented, the average LMP is the same for both ERC-SCUC and VI-ERC-SCUC cases. It is evident that the proposed VI-LRC-SCUC model provides better results than two other models. The average congestion LMP decreases from 29.09 $/MWh to 21.08 $/MWh when virtual inertia is included, and the average LMP decreases considerably from 45.89 $/MWh to 37.88 $/MWh. This implies that virtual inertia ancillary services can significantly improve the market efficiency when locational RoCoF constraints are included in the day-ahead scheduling SCUC model.

TABLE VII
MARKET RESULTS (AVERAGED OVER 24 HOURS) WITH DIFFERENT SCUC MODELS

| Model | Load payment [$/h] | Generator revenue [$/h] | Generator cost [$/h] | Generator rent [$/h] | Congestion revenue [$/h] |
|---|---|---|---|---|---|
| T-SCUC | 80,476 | 72,589 | 37,141 | 35,448 | 7,887 |
| ERC-SCUC | 82,975 | 59,651 | 41,189 | 18,462 | 23,324 |
| VI-ERC-SCUC | 83,202 | 62,005 | 40,475 | 21,530 | 21,197 |
| LRC-SCUC | 127,770 | 52,882 | 43,791 | 9,091 | 74,888 |
| VI-LRC-SCUC | 81,112 | 61,827 | 43,315 | 18,512 | 19,285 |

TABLE VII details the market results of all cases. The congestion revenue is relatively low for both T-SCUC and ERC-SCUC, comparing to the proposed LRC-SCUC model. This indicates that with location based RoCoF constraints included, network congestion contributes more to the differences in nodal LMPs and increases load payment. It is also evident that imposing virtual inertia leads to significant drop in the load payment for the proposed LRC-SCUC model, as well as the congestion revenue. Moreover, it is interesting to observe although VI-ERC-SCUC achieves a lower cost than ERC-SCUC, its load payment is even higher. In general, simulation results have shown that introducing virtual inertia can improve the flexibility of generators which might be strictly constrained by frequency related constraints, and the virtual inertia enabled generation flexibility may subsequently lower the burden of transmission lines and reduce the network congestion.

## VI. CONCLUSIONS

In this paper, the concept of locational frequency security is introduced. We first investigate the impact of Fiedler mode on locational frequency dynamics, and then the expression of locational frequency dynamics is defined accounting for *G*-1 contingency in multi-machine systems. To capture the highest locational RoCoF during the oscillation, a multiple-measurement-window method is introduced. Furthermore, a piecewise linearization based method is then proposed to convert the non-linear frequency constraints into linear frequency constraints in the proposed LRC-SCUC model, which allows us to optimally schedule the synchronous inertia as well as inertial services provided by non-synchronous resources to meet the minimum system inertia requirement for power systems with higher RES integration.

Simulation results show that imposing location based RoCoF constraints in the SCUC model can ensure the locational frequency security during worst-case contingency event. Such RoCoF-related constraints also significantly affect the scheduling of synchronous generators and consequently the expected system cost. The effect of virtual inertia on inertia pricing and market efficient is examined. Results imply that the proposed VI-LRC-SCUC model can improve the market efficiency. Compared to VI-ERC-SCUC model, VI-LRC-SCUCU is more sensitive to inertia price. As a result, imposing virtual inertia techniques can largely reduce the total cost by avoiding unnecessary commitment of extra expensive synchronous generators while meeting the system inertia requirement.